# From Untamed Black Box to Interpretable Pedagogical Orchestration: The Ensemble of Specialized LLMs Architecture for Adaptive Tutoring

Nizam Kadir[1][0000-0002-6725-1133]

[1] Singapore University of Technology and Design, Singapore 487372, Singapore
`nizam_kadir@mymail.sutd.edu.sg`

**Abstract.** Monolithic Large Language Models (LLMs) used in educational dialogue often behave as "black boxes," where pedagogical decisions are implicit and difficult to audit, frequently violating instructional constraints by providing answers too early. We introduce the Ensemble of Specialized LLMs (ES-LLMs) architecture that separates decision-making from wording. Pedagogical actions are selected by a deterministic rules-based orchestrator coordinating specialized agents—covering tutoring, assessment, feedback, scaffolding, motivation and ethics—guided by an interpretable Bayesian Knowledge Tracing (BKT) student model. An LLM renderer surface-realizes the chosen action in natural language. This design emphasizes reliability and controllability: constraints such as "attempt-before-hint" and hint caps are enforced as explicit rules, and the system logs per-turn agent traces and constraint checks. Validation of pedagogical quality via human expert reviewers ($N$=6) and a multi-LLM-as-Judge panel (six state-of-the-art models) showed that ES-LLMs were preferred in 91.7% and 79.2% of cases, respectively. The architecture significantly outperformed monolithic baselines across all seven dimensions, particularly in Scaffolding & Guidance, and Trust & Explainability. Furthermore, a Monte Carlo simulation ($N$=2,400) exposed a "Mastery Gain Paradox," where monolithic tutors inflated short-term performance through over-assistance. In contrast, ES-LLMs achieved 100% adherence to pedagogical constraints (e.g., attempt-before-hint) and a 3.3x increase in hint efficiency. Operationally, ES-LLMs reduced costs by 54% and latency by 22% by utilizing stateless prompts. We conclude that structural decoupling is essential for transforming stochastic models into trustworthy, verifiable and resource-efficient pedagogical agents.

## 1 Introduction

Large Language Models (LLMs) have solved the *"Fluency Problem"* in educational dialogue but exacerbated the *"Control Problem"* [1]. In safety-critical tutoring, monolithic tutors often violate pedagogical constraints—giving away answers too early—due to the inherent bias of general-purpose LLMs toward frictionless user satisfaction [2]. We term this the **Mastery Gain Paradox**: a phenomenon where users achieve high short-term performance metrics (via over-assistance) while their latent mastery stagnates or degrades.

Classical Intelligent Tutoring Systems (ITS) prioritize theoretically grounded decision logic (e.g., Bayesian Knowledge Tracing (BKT) [3]) but lack generative flexibility. To address this gap, this paper investigates the following central research problem: How can we harness the conversational fluency of LLMs for adaptive tutoring while strictly guaranteeing adherence to pedagogical constraints and procedural fairness? We propose **Ensemble of Specialized LLMs (ES-LLMs)**, a hybrid neuro-symbolic architecture that treats the tutor not as a monolithic tool, but as a *team of coordinated specialists*. Importantly, the term "ensemble" here refers to the systematic coordination of multiple deterministic, rule-based pedagogical policies (e.g., assessment, scaffolding, ethics) computing in parallel, rather than an ensemble of expensive stochastic LLM calls. By decoupling the *deterministic pedagogical decision* from the single-call LLM *surface realization*, ES-LLMs restores auditability and trust [1].

## 2      Related Work

### 2.1      From Rule-Based ITS to Generative Tutors

Classical ITS like Cognitive Tutors [4, 5] use model tracing and BKT [3] to deliver personalized, rule-based instruction with proven learning gains [6, 7]. Performance Factors Analysis (PFA) [8] and neural alternatives exist but lack the transparency needed for real-time safety orchestration. Recent work explores fine-tuning LLMs for feedback [9] and teachable agents [10]. However, even optimized monolithic models struggle with negative constraints (e.g., "do not reveal the answer"). Multi-agent approaches suggest decomposing roles but often leave the logic implicit in prompts. ES-LLMs makes these roles deterministic and rule-governed.

### 2.2      Simulation and Automated Evaluation Methods

Simulation-Based Mastery Learning (SBML) [11] posits that mastery in simulation (T1) is a prerequisite for translational outcomes in real-world practice (T2) and downstream impact (T3). We apply this framework by utilizing Monte Carlo simulations with synthetic students to stress-test reliability as validation of pedagogical rigor. Traditional metrics like BLEU are poor for tutoring. Emerging "LLM-as-judge" methods [12] and Panels of LLMs (PoLL) [13] offer scalability, but human expert review remains the gold standard.

### 2.3      Procedural Fairness and Interpretability

Fairness in AIED [14] requires more than demographic parity; it necessitates process consistency (procedural fairness), which interprets as architectural safety in our framework. It is also worth noting that fairness ties into the interpretability of the tutor: an interpretable system allows stakeholders to detect if it is treating some students differently [15]. By maintaining a rule-based decision layer, educators can audit and adjust the policies to be fair (for instance, by adding rules that ensure all students receive motivational feedback after a certain number of errors, to avoid neglecting quieter students). As we progress to real-world deployment, we plan to incorporate privacy-preserving collection of demographic data (with consent) to directly evaluate fairness metrics and mitigate any biases. For this paper's scope, we demonstrate how fairness is considered via design (agents like EthicsBot) and via offline cohort evaluations.

## 3 System Architecture and Implementation

Fig. 1 illustrates the architecture of our **Ensemble of Specialized LLMs (ES-LLMs)** tutoring system. The design is a pipeline with multiple stages, integrating data-driven modeling and rule-based decision making.

Conceptually, ES-LLMs can be mapped to a triarchic tutoring blueprint—Expert Model (domain/solution knowledge), Learner Model (BKT mastery), and Tutor Model (agent policies and orchestrator)—with the LLM restricted to surface realization.

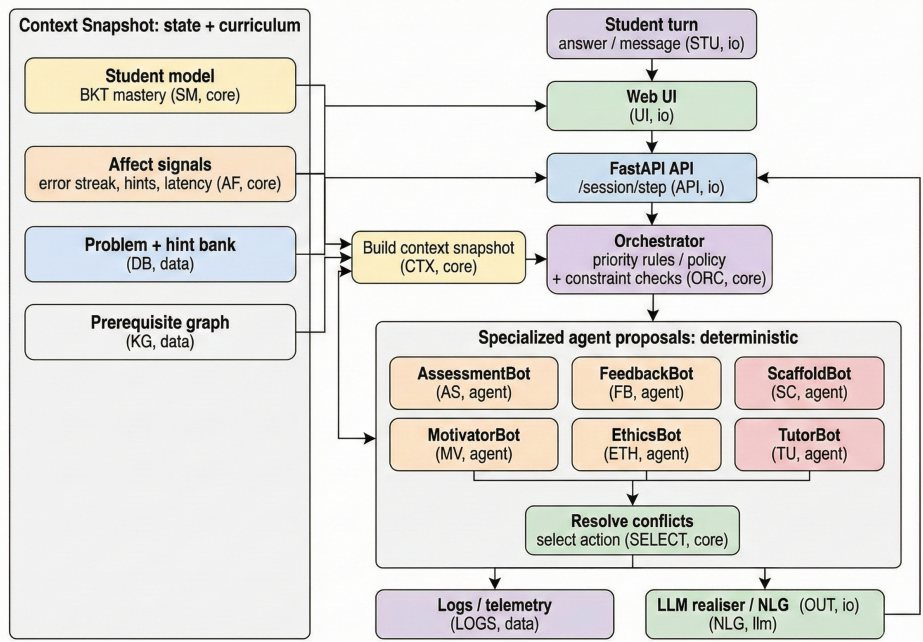

**Fig. 1.** Architecture of the ES-LLMs adaptive tutoring system. A BKT student model produces mastery posteriors that drive specialized agent proposals, while a deterministic rules-based orchestrator enforces sequencing and constraints (e.g., attempt-before-hint and hint caps) and records per-turn traces. An LLM renderer is used for wording of the selected action; all decisions and constraint checks are logged for system-level evaluation.

### 3.1    Data Ingestion and Feature Engineering

The system ingests student interaction logs from ASSISTments 2017 [16] tutoring interactions ($N$=942,816) and computes a set of pedagogically relevant features for each attempt. These features include rolling accuracy (the student's recent success rate on a skill), hint usage counts and time since last hint, time-on-task durations, number of opportunities practiced for each skill, and other indicators of possible frustration (such as "wheel-spinning" when many attempts are made without success). The features are updated after each student action and stored for use by the student model. This preprocessing ensures that the tutor has a rich, contextual state representation at each turn.

### 3.2    Student Modeling

The deployed tutor uses per-skill BKT as its runtime student model [3]. BKT maintains a mastery posterior for each skill ($N$=102) and updates this belief after each student attempt (e.g., correct/incorrect, with optional slip/guess and learning parameters). Agents and the orchestrator consult these mastery estimates to decide when to provide feedback, hints, scaffolding, or progression.

### 3.3    Pedagogical Agents: An Ensemble of Specialized LLMs

The Orchestrator coordinates six core specialized agents (detailed in Table 1a) using a strict priority hierarchy: **Safety First** (EthicsBot filters harmful content), **Assessment Second** (AssessmentBot tracks mastery state), and **Pedagogy Third** (ScaffoldBot and MotivatorBot provide support). To ensure reliability, the ensemble operates in two stages. First, a **deterministic decision layer** evaluates the learner's state using strict rules and knowledge tracing (BKT) to select the single active agent, effectively muting the others. This modular design ensures that pedagogical strategy (e.g., scaffolding) is never overridden by generative fluency. Second, the selected agent is instantiated via a single, tailored LLM call. By structurally isolating each agent's logic before generation, the architecture creates a 'mixture of prompts' capable of specialized expertise without the latency or hallucination risks of uncontrolled agent interaction.

Table 1b summarizes the supporting architecture modules in the ES-LLMs pipeline, including the meta-orchestrator, pedagogical policy, domain expert, affect detector, and LLM renderer, linking each component to its operational function within the system.

**Table 1a.** Core Pedagogical Agents in ES-LLMs, their theoretical foundation, and implementation rules

| Agent | Role | Evidence Basis | Encoding in System |
|---|---|---|---|
| **AssessmentBot** | Mastery estimation | Knowledge tracing / BKT [3, 17, 18], assessing learning [19] | Updates after each attempt |
| **FeedbackBot** | Corrective feedback | Feedback effectiveness [21, 22] | Emits CONFIRM/ NUDGE/ REMEDIATE from correctness and mastery |
| **ScaffoldBot** | Hinting/ scaffolding | Scaffolding in tutoring [23, 24] | Emits HINT_ MIN/ MED/ FULL from error windows and hint caps |
| **MotivatorBot** | Affective support | Motivation and affect dynamics [25, 26] | Encouragement for low confidence or error streaks |
| **EthicsBot** | Guardrails | Help-seeking and gaming constraints [27, 28] | Enforces attempt- before- hint and max- hint rules |
| **TutorBot** | Progression/ next item | Mastery learning and time-to-learn [29, 30, 31] | Chooses NEXT when mastery exceeds threshold |

**Table 1b.** Supporting Architecture Modules and their roles in the system pipeline

| *Modules* | *Role* | *Evidence basis* | *Encoding in system* |
|---|---|---|---|
| **Meta-Orchestrator** | Decision coordination | ITS architectures and effectiveness [5, 6, 7] | Coordinates modules, orders actions, aggregates into one response |
| **Pedagogical Policy** | Action sequencing | Rule-based ITS policies [5, 20] | Orders agents by correctness, mastery, affect, streaks |
| **Affect Detector** | Affect state | Affective dynamics in learning [26] | Heuristic affect inference from error patterns |
| **Domain Expert Client** | Content hints | Human tutoring explanations [23, 32] | Short content hint/explain injected into feedback/scaffold |
| **Renderer (LLM)** | Wording and aggregation | Feedback should be specific and concise [21, 22] | One LLM call to surface-realize aggregated decisions |

### 3.4 The Orchestrator

The Orchestrator (or Meta-Orchestrator) functions as a deterministic, rule-based policy manager that arbitrates candidate actions from all agents to determine execution order.

We implement a priority hierarchy based on **Subsumption Architecture** [33], a robotics design pattern that manages competing goals through hierarchical suppression. In this educational adaptation, higher-priority layers suppress lower-priority ones to ensure pedagogical safety: **Safety (EthicsBot) > Assessment > Feedback > Scaffolding > Motivation**. For instance, if EthicsBot detects a constraint violation (e.g., student attempt count = 0), it outputs a DENY_HINT action that immediately suppresses ScaffoldBot's HINT_FULL proposal. Conversely, if no safety rules are violated and the student's BKT mastery probability exceeds a threshold ($pL > 0.95$), TutorBot triggers a NEXT_PROBLEM action. The Orchestrator's output is a structured representation of the final tutor turn, passed as an ordered list of decisions to a single aggregation step for surface realization.

### 3.5   LLM-based Renderer

The system invokes an LLM to generate the dialogue text for the tutor's turn. Instead of one LLM handling everything end-to-end, we use a single aggregation prompt that combines the ordered agent decisions into one concise response. In our prototype, the default realizer configuration uses gpt-4o-mini with temperature 0.3 and max_tokens=120 for short, stable surface realizations; these parameters are part of the runtime configuration and are logged per run. The realizer prompt is structured as (i) a system role instruction ("compose a single response from the ordered decisions; do not change the decisions") and (ii) a user message containing the ordered decisions plus a compact, sanitized context (skill, mastery posterior, attempt counts, constraint state, and relevant dialogue history). The key is that the LLM *does not decide* what pedagogical action to take; it only phrases the chosen actions in one response. If an API key is unavailable, the system falls back to deterministic wording templates.

### 3.6   Delivery and Interface

The finalized tutor message (or sequence of messages) is delivered to the student. In our implementation, we built a simple chat-style web interface where the conversation between the student and the tutor (ES-LLMs or baseline) is displayed. Each tutor turn includes badges indicating which agents were activated (for the researchers' debugging, this would be hidden from actual students). The system also logs all interactions and model outputs for analysis.

**User Interface Design.** Fig. 2 shows the web-based tutoring interface through which students interact with the ES-LLMs system. Unlike black-box LLM tutors where decision-making is opaque, the ES-LLMs interface makes the system's pedagogical reasoning visible to researchers, educators, and learners. The interface follows established design principles for intelligent tutoring systems, presenting a clean, distraction-free environment that focuses learner attention on the mathematical problem at hand. The layout consists of three primary zones: (1) a problem presentation area displaying the current mathematics question, (2) a student input field with a confidence slider (1-5 scale) where learners enter their answers, and (3) a tutor response area where the system provides feedback, scaffolding, or encouragement based on the orchestrator's pedagogical decisions.

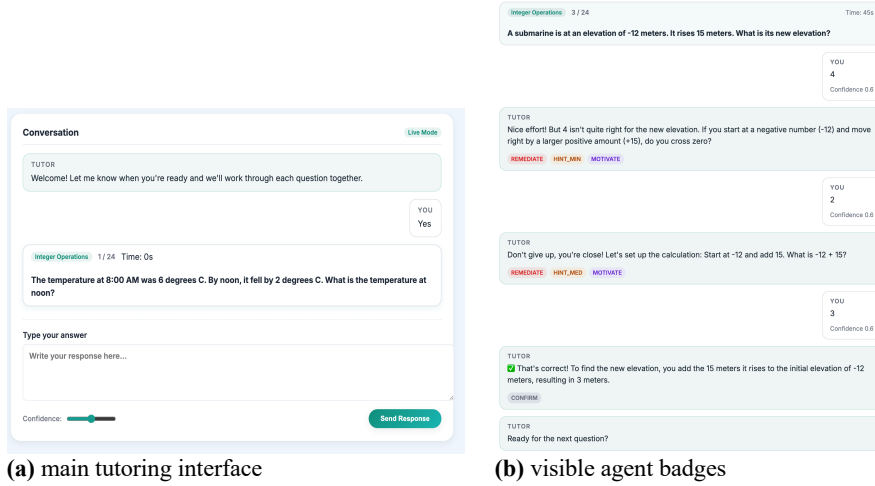

**(a)** main tutoring interface    **(b)** visible agent badges

**Fig. 2.** The ES-LLMs interface: **(a)** student input and **(b)** interpretable agent orchestration

## 4 Methodology and Experimental Design

### 4.1 Research Questions

We compare ES-LLMs to a monolithic baseline across two primary dimensions: (1) Does decoupling decision-making from generation resolve the "Mastery Gain Paradox" by improving pedagogical constraint adherence and hint efficiency? **(**2) Quality and Operational Efficiency: How does ES-LLMs compare to monolithic tutors in expert-rated pedagogical quality (across seven dimensions) and in computational resource efficiency (latency and token cost)?

### 4.2 Baseline (Direct LLM)

The baseline is a single-call, monolithic tutor that uses the exact same LLM (gpt-4o-mini) and decoding settings as ES-LLMs' renderer and agent calls. We intentionally selected a lightweight, standard instruction-tuned model (gpt-4o-mini) for both systems to isolate the impact of our orchestrator architecture from the inherent advanced reasoning capabilities of frontier models (e.g., o1). The baseline was allowed a higher max_tokens limit of 300 to ensure its verbosity was not artificially truncated. It receives the same problem context and dialogue history but makes and phrases pedagogical decisions end-to-end in a single prompt.

### 4.3 Tutoring Scenarios and Dialogue Generation

To reflect authentic learner behaviors, we used the ASSISTments 2017 dataset [16] (students $N$=1,709) in $k$-means clustering across three dimensions (skill difficulty, interaction sequence, and learner profile). We identified 8 distinct interaction signatures (e.g. Clean Correct, Hint Abuse, Deep Struggle). We crossed this with 3 difficulty tiers to construct a controlled testbed of 24 fixed scenarios anchored to the skills in the Common Core State Standards for Mathematics (CC-SSM) for middle school (typically ages 11-14).

### 4.4    Monte Carlo Simulation with Synthetic Students

We conducted a large-scale **Monte Carlo simulation** ($N$=2,400). We instantiated **Synthetic Student Agents** based on BKT parameters ($P(L_0)$, $P(T)$, $P(S)$, $P(G)$) derived from the ASSISTments 2017 dataset [16]. We defined four learner archetypes (Struggling, Low, Average, High) using $k$-means clustering and ran 600 stochastic simulations per archetype. In each run, student parameters were perturbed by Gaussian noise to ensure robustness. This setup allows us to measure **Procedural Fairness**; verifying that the system's treatment (constraints) remains consistent despite stochastic variations in student behavior, and to statistically compare the learning efficiency of ES-LLMs against the baseline. We prioritized computational metrics from these 2,400 runs to evaluate the system's structural integrity. Drawing on the framework of Simulation-Based Mastery Learning (SBML) [11], we utilized synthetic student agents to ensure that the tutor's pedagogical policy holds under stochastic variation. The simulation serves as a T1 validation of the 'Deliberate Practice' conditions required for effective learning.

### 4.5    Pedagogical Quality Evaluation by Human Experts and LLM Judges

Human experts consisting of educators and educational researchers ($N$=6) evaluated the paired tutoring dialogues. We utilized a rigorous, double-blind 5-point Likert rubric adapted to assess seven critical pedagogical dimensions: Adaptivity, Scaffolding & Guidance, Ethical Reasoning, Engagement, Feedback Quality, Tone & Style, and Trust & Explainability. As a secondary signal, we conducted multi-LLMs-as-Judge evaluations using six state-of-the-art LLMs (Gemini 3 Pro, Qwen3-Max, Kimi K2, DeepSeek-V3, Grok-2, ChatGPT-5.2). Each model was prompted to act as an expert educator and assessed the 24 paired interactions using the exact same rubric applied by human evaluators, capturing both quantitative scores and qualitative reasoning.

## 5    Results

### 5.1    Computational Metrics: The Efficiency-Fidelity Trade-off

Monte Carlo simulations (Table 2) reveal key differences across pedagogical fidelity and resource efficiency dimensions. **The Mastery Gain Paradox and Constraint Adherence.** The simulations reveal a diagnostic failure we label the **Mastery Gain Paradox**: the baseline achieved a higher mastery gain ($\Delta$ = +0.70 vs. +0.40), but only by providing **12x more hints** (6.8 vs. 0.44 per dialogue). The baseline defaulted to giving answers early to minimize friction, whereas ES-LLMs achieved **100% adherence** to the attempt-before-hint constraint, encouraging productive struggle [7]. In addition, ES-LLMs has higher hint efficiency (which is calculated as the mastery gain per hint).

**Table 2.** Computational Metrics from Monte Carlo Simulation ($N$=2,400). Significance via Wilcoxon Signed-Rank Test.

| Metric Type | Baseline | σ | ES-LLMs | σ | Wilcoxon $p$ |
|---|---|---|---|---|---|
| Mastery Gain (BKT) | 0.70 | 0.19 | 0.40 | 0.42 | < 0.001*** |
| Constraint Adherence | 62.4% | 22.1 | 100.0% | 0.0 | < 0.001*** |
| Hint Efficiency | 0.10 | 0.05 | 0.33 | 0.38 | <0.001*** |
| Latency (ms) | 800 | 150 | 625 | 110 | <0.05* |
| Tokens | 1300 | 220 | 590 | 95 | <0.001*** |

**Resource Efficiency.** Beyond pedagogical gains, ES-LLMs demonstrates substantial computational advantages. By separating deterministic decision logic from natural language generation, the architecture achieves **54% cost reduction** (590 vs. 1,300 tokens per turn, p<0.001) and **22% latency improvement** (625ms vs. 800ms, p<0.05). The efficiency stems from stateless NLG prompts—agents pass only decision-relevant context (action type, skill state, mastery level) rather than full conversation history, reducing input tokens 2.2×. At scale, this translates to **$1.41 savings per 1,000 sessions** while maintaining 100% constraint adherence, demonstrating that pedagogical correctness and operational efficiency are complementary, not competing, design goals.

### 5.2    Unanimous Superiority in Pedagogical Quality

Six human experts and six LLM judges evaluated the 24 scenario pairs. Out of the 144 total comparative evaluations (6 x 24 items), the ES-LLMs architecture was selected as superior in **91.7%** ($n$=132) and **79.2%** of cases ($n$=114) by humans and LLM judges respectively. This is compared to 6.9% ($n$=10) and 13.2% for the baseline ($n$=19) respectively. Tied outcomes were 1.4% ($n$=2) and 7.6% ($n$=11) respectively. Table 3 shows aggregate dimensional scores. ES-LLMs significantly outperformed baseline on all 7 dimensions for both human experts and LLM judges.

**Qualitative Themes.** Thematic analysis of evaluator reasoning—from both human experts and LLM judges—revealed three critical pedagogical distinctions:

*Prevention of "Gaming the System".* ES-LLMs consistently detected and mitigated hint abuse (e.g., students spamming "idk" or clicking hints without reading), withholding answers until effort was demonstrated. Human experts noted this in 75% of cases ($n$=18), with one educator commenting: *"Dialogue A [ES-LLMs] correctly detected 'idk' spam and withheld the answer"*. The baseline provided answers immediately, enabling gaming strategies.

*Deep Remediation vs. Surface Loops.* ES-LLMs employed deeper remediation strategies (using analogies like money vs temperature), while the baseline tended to repeat the same formulaic advice. Human experts highlighted this in 58% of evaluations ($n$=14), noting: *"Dialogue A [ES-LLMs] changed strategies with a concrete pizza* *analogy when the student was deeply stuck"*. LLM judges concurred, with three models (DeepSeek-V3, Kimi K2, ChatGPT-5.2) rating ES-LLMs higher on Scaffolding & Guidance (mean difference: +1.99 human, +1.84 LLM, *p*<.001).

**Table 3.** Human Expert and Multi-LLMs-as-Judge Aggregate Scores comparing ES-LLMs and Baseline. Ratings are on a 1–5 Likert scale across 24 tutoring dialogue pairs. Positive differences indicate ES-LLMs superiority. All differences are statistically significant (*p*<.01 for Ethical Reasoning and Tone and Style; *p*<.001 for all other dimensions).

| Dimension | ES-LLMs Mean | | Baseline Mean | | Difference | |
|---|---|---|---|---|---|---|
| | LLMs | Humans | LLMs | Humans | LLMs | Humans |
| Adaptivity | 4.82 | 4.62 | 3.55 | 2.96 | +1.27 | +1.66 |
| Scaffolding & Guidance | 4.78 | 4.67 | 2.94 | 2.68 | +1.84 | +1.99 |
| Ethical Reasoning | 4.95 | 4.88 | 4.10 | 4.01 | +0.85 | +0.87 |
| Engagement | 4.86 | 4.29 | 3.42 | 3.12 | +1.44 | +1.17 |
| Feedback Quality | 4.89 | 4.67 | 3.78 | 3.59 | +1.11 | +1.08 |
| Tone and Style | 5.00 | 4.76 | 4.20 | 4.15 | +0.80 | +0.61 |
| Trust & Explainability | 4.70 | 4.67 | 3.10 | 2.80 | +1.60 | +1.87 |

*Ethical Reasoning.* Both evaluator types rated systems highly on safety, but human experts emphasized ES-LLMs' avoidance of misleading feedback in 46% of cases (*n*=11): *"Dialogue B [baseline] said the answer was 'close' when completely wrong. Dialogue A [ES-LLMs] never misled"*. Despite strong performance by both systems (ES-LLMs: 4.88, Baseline: 4.01), ES-LLMs maintained an edge (Δ=+0.87, *p*<.01).

**Efficiency Trade-off.** In a minority of cases where the baseline outperformed ES-LLMs (e.g., 3+0, 1+1) human experts noted the orchestration overhead: *"A difficulty assessment layer could have bypassed unnecessary scaffolding."* Three LLM judges also flagged these edge cases, suggesting future work should include adaptive orchestration depth.

### 5.3   Interpretable Constraint Enforcement: From Opaque to Transparent

Fig. 4a shows how the interface transforms invisible algorithmic decisions into transparent feedback. When a student triggers a gaming detector, the system displays a *"HINT DENIED"* badge with an explicit pedagogical rationale (*"I need to see you try first..."*). This visualizes the system's deterministic logic, converting a potential "blackbox" refusal into an interpretable educational moment that reinforces the *attempt-before-hint* protocol.

**The "Team Player" Ensemble in Action.** A 6-turn deep remediation sequence illustrates the orchestrator's strategic escalation of support (see Fig. 4b). The progression of agent badges (from *REMEDIATE DEEP* to *HINT FULL*) reveals how the system dynamically shifts specialists in response to persistent struggle. The dialogue also highlights the interleaving of *MotivatorBot's* encouragement alongside *ScaffoldBot's* instructional guidance. By explicitly labeling which agent contributes to each turn, the interface demonstrates the collaborative nature of the ensemble, providing a level of transparency and context-awareness that opaque monolithic models lack.

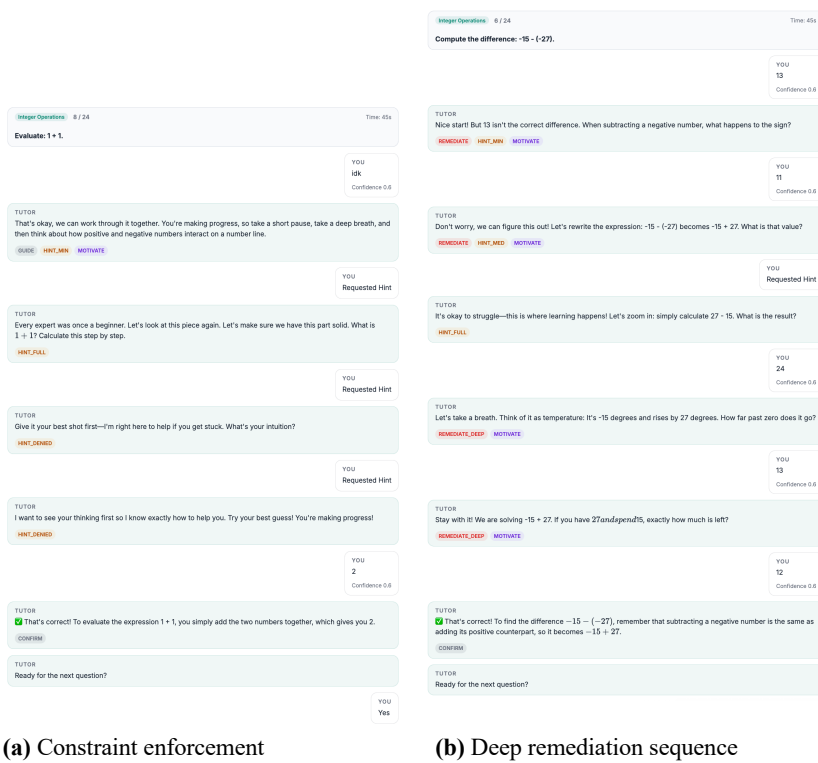

(a) Constraint enforcement     (b) Deep remediation sequence

**Fig. 4.** Transparency in action: **(a)** HINT_DENIED rationale and **(b)** agent escalation traces

## 6 Discussion

Our architecture targets a reliability gap in LLM tutoring: monolithic tutors can be prompt-sensitive and difficult to constrain, while educational settings often require consistent enforcement of pedagogical rules. Below we discuss implications and limitations of an orchestrated, multi-agent ES-LLMs design.

### 6.1 Implications for Adaptive Tutor Design

A core implication of ES-LLMs is that pedagogical policy can be externalized as explicit, testable rules. This allows for the rigorous enforcement of 'minimum passing standards' and 'deliberate practice' protocols, which are identified as the essential components of simulation-based mastery learning [11]. Unlike monolithic models where these standards are implicit and probabilistic, ES-LLMs guarantees the rigorous practice environment necessary for translational outcomes.

Another implication is the maintainability and extensibility of such a system. The modular ES-LLMs design means one can improve or swap out components independently for example, replacing the BKT student model with a more advanced model (e.g., deep knowledge tracing) without changing the orchestration or wording layers. In contrast, the monolithic approach is a single entity: if it is not performing well, the primary recourse is prompt engineering or fine-tuning, which are less transparent than adjusting an explicit rule or threshold in a multi-agent system. This modular approach aligns with good software engineering (and MLOps) principles. In production, a multi-agent tutor may also be easier to troubleshoot: if responses are off, one can inspect whether the issue was with the student model estimate, a bot's rule, or the LLM phrasing; with a monolith, it is often unclear why a decision was made.

### 6.2    Broader Implications: The Policy-Generation Decoupling Paradigm

Beyond intelligent tutoring, the ES-LLMs architecture illustrates a generalizable solution for high-stakes AI applications: the **structural decoupling of decision-making (policy) from surface realization (generation)**. This pattern offers a robust alternative to purely probabilistic models in domains like **healthcare** and **customer operations**, where strict adherence to clinical protocols or escalation logic is non-negotiable. By externalizing control logic into a deterministic orchestrator, the system replaces malleable prompt-based guardrails with architectural constraints. This approach successfully reconciles the tension between the necessary rigidity of rule-based compliance and the conversational fluency of Large Language Models, providing a template for trustworthy AI alignment.

### 6.3    Alignment Between AI and Human Evaluation

One interesting next step is the concept of **calibrating LLM evaluators** for educational criteria. Perhaps an LLM could be fine-tuned on a dataset of human ratings of tutor responses to become a more accurate "education-specific judge." There is emerging work on improving LLM-evaluator prompts and calibration for better human alignment [12], and our study can contribute data to that effort in the tutoring domain. In practice, a combination of human and LLM evaluation may yield the best efficiency and reliability: use LLM judges to triage obvious failures and monitor regressions but use periodic human audits to ensure pedagogical quality is truly high [12].

### 6.4    Fairness and Ethical Auditability

We define fairness in AI tutoring not merely as demographic equity, but as **behavioral consistency**—ensuring uniform pedagogical support across varying skill levels [14]. The ES-LLMs architecture achieves **Safety by Design** by offloading constraint enforcement to a deterministic orchestrator. This eliminates the "stochastic unfairness" inherent in monolithic LLMs, where identical student inputs can yield varying levels of support due to model probability. Our simulations confirmed that ES-LLMs maintained 100% protocol compliance across all behavioral cohorts (e.g., high-mastery vs. gaming behaviors), whereas the baseline exhibited high variance that could disadvantage struggling learners. Furthermore, unlike "black box" tutors, the interpretable decision layer allows for granular auditing of policy enforcement, providing a necessary foundation for future demographic equity studies in real-world deployments.

### 6.5    Limitations

Several limitations must be acknowledged. First, and primarily, we report quantitative efficacy metrics based on Monte Carlo synthetic students rather than real-world learning outcomes. While these simulations successfully establish architectural reliability (e.g., verifying 100% constraint adherence and procedural fairness under massive stochastic variation), synthetic learners governed by idealized BKT parameters do not perfectly mirror the messy cognitive and affective realities of human classrooms—such as unpredictable slips or attention drift. As emphasized by Simulation-Based Mastery Learning frameworks [11], this computational validation constitutes a necessary Phase I trial to ensure system safety before deployment. Controlled, in-classroom studies with human learners are required to establish true educational effectiveness.

Second, our evaluation scope was restricted to foundational, single-problem mathematics tutoring. Findings may not generalize to open-ended domains or multi-session interactions, and future applications must assess the architecture's capacity to orchestrate scaffolding for semantically complex, multi-step word problems. Third, we compared ES-LLMs against a standard prompt-only baseline; incorporating stronger baselines tailored with few-shot or self-critique methods might alter the comparative results. Finally, our multi-agent architecture is inherently more complex than a monolithic solution, introducing potential points of failure and requiring more engineering resources to maintain.

### 6.6    Future Work

Future research will prioritize a user study with real students to validate learning outcomes and engagement. In the taxonomy of Simulation-based Mastery Learning (SBML) [11], the current study establishes strong T1 outcomes (simulation-based reliability and constraint adherence). The next phase of research aims to establish T2 translational outcomes (improved problem-solving practices in the classroom) and ultimately T3 outcomes (long-term mastery retention). The simulation results provide the necessary confidence that the system is safe and rigorous enough for this transition.

We also plan to explore **adaptive orchestration policies** that move beyond static rules, potentially using reinforcement learning to fine-tune agent sequencing. Since the baseline outperformed ES-LLMs on trivial problems, we propose adding a **difficulty assessment layer**. This would route simple items directly to the LLM for immediate confirmation while reserving full orchestration for challenging tasks, improving efficiency without sacrificing rigor. Improving the **LLM phrasing layer** is another key goal. We intend to fine-tune models on high-quality tutor phrasing to enhance tone and brevity while ensuring the LLM remains strictly a wording engine.

## 7    Conclusion

We introduced ES-LLMs, an architecture that decouples deterministic pedagogical decisions from generative wording. This structural separation ensures auditable tutoring, addressing the reliability risks of monolithic models. Human and LLM-as-Judge evaluations confirmed that ES-LLMs significantly outperform baselines in pedagogical quality. Furthermore, simulations ($N$=2,400) exposed a "Mastery Gain Paradox" where monolithic tutors inflate performance through over-assistance. In contrast, ES-LLMs achieved 100% constraint adherence and a 3.3x increase in hint efficiency. Operationally, the architecture reduced costs by 54% and latency by 22%. Future work will translate these simulation-based gains to controlled studies with human learners.

**Acknowledgments.** We would like to thank Assoc. Prof. Dorien Herremans, Dr Nachamma Sockalingam, Mr. Yang Xingyu and Mr. Syed Adeel Ahmed Mustaq Ahamed for their involvement in this work. We also extend our deepest gratitude to the six human expert reviewers for their time and essential insights in evaluating the pedagogical quality of the system.

**Disclosure of Interests / Ethics Approval.** The author has no competing interests to declare that are relevant to the content of this article. This study was approved by the Singapore University of Technology and Design (SUTD) Institutional Review Board under reference code S-25-782.

**Data Availability and Reproducibility.** The pedagogical ruleset, evaluation rubrics, and sample scenario data are open-source and available at: https://github.com/nizamkadirteach/aied2026-es-llms.